\documentclass[manuscript,screen]{acmart}

\AtBeginDocument{%
  \providecommand\BibTeX{{%
    \normalfont B\kern-0.5em{\scshape i\kern-0.25em b}\kern-0.8em\TeX}}}

\setcopyright{acmcopyright}
\copyrightyear{2020}
\acmYear{2020}
\acmDOI{10.1145/1122445.xxxxxxx}

\acmConference[DeepSpatial 2020]{DeepSpatial 2020: ACM SIGKDD Workshop on Deep Learning for Spatiotemporal Data, Applications, and Systems}{August 24, 2020}{San Diego, CA}
\acmBooktitle{DeepSpatial 2020: ACM SIGKDD Workshop on Deep Learning for Spatiotemporal Data, Applications, and Systems, August 24, 2020, San Diego, CA}
\acmPrice{15.00}
\acmISBN{978-1-4503-XXXX-X/18/06}

\usepackage{tikz}
\usepackage{pgfplots}
\usepackage{pgfplotstable}
\usepackage{tikz}
\usetikzlibrary{positioning}
\usetikzlibrary{trees}
\usetikzlibrary{plotmarks}
\usepackage{tikz}
\usepackage{subfigure}
\usetikzlibrary{patterns}
\usepackage{mathtools}
\pgfplotsset{compat=1.14}

\begin{document}

\title{DeepSampling: Selectivity Estimation with Predicted Error and Response Time}
\titlenote{This work is supported in part by the National Science Foundation (NSF) under grants IIS-1838222 and CNS-1924694}

\author{Tin Vu}
\orcid{}
\affiliation{%
  \institution{Computer Science and Engineering}
  \institution{University of California, Riverside}
}
\email{tin.vu@email.ucr.edu}

\author{Ahmed Eldawy}
\orcid{}
\affiliation{%
  \institution{Computer Science and Engineering}
  \institution{University of California, Riverside}
}
\email{eldawy@ucr.edu}

\renewcommand{\shortauthors}{Tin Vu and Ahmed Eldawy}

\sloppy
\begin{abstract}
The rapid growth of spatial data urges the research community to find efficient processing techniques for interactive queries on large volumes of data. Approximate Query Processing (AQP) is the most prominent technique that can provide real-time answer for ad-hoc queries based on a random sample. Unfortunately, existing AQP methods provide an answer without providing any accuracy metrics due to the complex relationship between the sample size, the query parameters, the data distribution, and the result accuracy.
This paper proposes DeepSampling, a deep-learning-based model that predicts the accuracy of a sample-based AQP algorithm, specially selectivity estimation, given the sample size, the input distribution, and query parameters. The model can also be reversed to measure the sample size that would produce a desired accuracy. DeepSampling is the first system that provides a reliable tool for existing spatial databases to control the accuracy of AQP.
\end{abstract}






\maketitle

\section{Introduction}
\label{sec:introduction}

Recently, there has been a notable increase in the amounts of spatial data collected by satellites, social networks, and autonomous vehicles. The main method that data scientists use to process this data is through {\em interactive exploratory queries}; i.e., an ad-hoc query that should be answered in a fraction of a second. Existing studies show that a response time of more than a few seconds to these queries would negatively impact the productivity of the users~\cite{LH14}. Unfortunately, existing big-spatial data systems~\cite{eldawy2015spatial,yu2015geospark,XLY+16, BVK+17,TYM+16}, require way more than that to run even the simplest queries, hence, they cannot answer interactive exploratory queries.

\begin{figure}[t]
    \centering
    \includegraphics[width=0.8\linewidth]{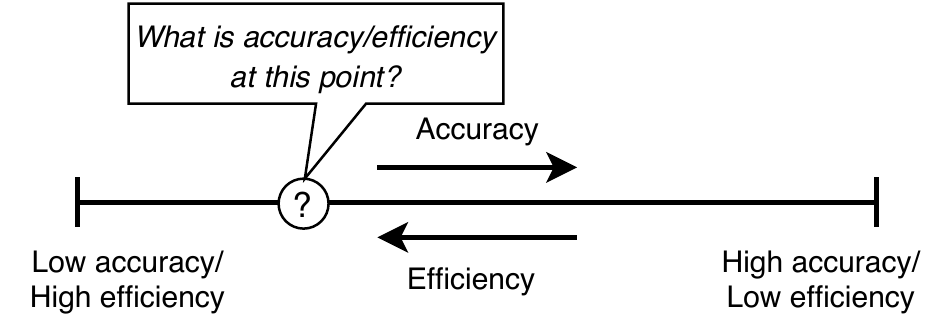}
    \caption{Trade-off between accuracy and efficiency in AQP}
    \vspace{-10pt}
    \label{fig:trade_off}
\end{figure}

The most viable solution to the interactive exploration problem is approximate query processing (AQP) which uses a small data synopsis, e.g., a sample, to provide an approximate answer within a fraction of a second. This technique provides up-to three orders of magnitude speedup with a very high accuracy for several fundamental problems, including selectivity estimation, clustering, and spatial partitioning~\cite{siddique2019comparing}. Figure~\ref{fig:trade_off} depicts the trade-off between the {\em accuracy} of the approximate answer and the {\em efficiency}, i.e., running time, which is highly correlated with the sample size. Unfortunately, this accuracy/efficiency trade-off is very hard to calculate which discourages many users from using AQP systems. Existing solutions either provide answers without any performance guarantee or make unrealistic assumptions such as uniform distribution or independence between dimensions~\cite{siddique2019comparing,CE17,siddique2019euler++,siddique2018experimental,JAS00,AN00,OR93,agarwal2013blinkdb,LNS90,PIH+96,I93}. This problem is particularly challenging due to the intertwined relationship between the sample size, query parameters, algorithm logic, data distribution, and result accuracy.

This paper proposes DeepSampling, a novel deep learning based model to predict the relationship between accuracy and relative sample size for AQP. The main challenge is how to build a model that works well for any spatial data distribution and query parameters. To solve this problem, we build a deep neural network that takes as input the query parameters and a histogram that represents the data distribution. This idea can work in two modes: \emph{1) given a sample size, it estimates the expected accuracy, or 2) given a desired accuracy, it calculates the required sample size.} The idea is generic and can work with any approximate algorithm by building a separate model for each one. DeepSampling can be integrated into any existing spatial data system that supports AQP. To the best of our knowledge, DeepSampling is the first system that supports predictable error AQP for spatial data analysis problems. We run an experimental evaluation on both synthetic and real data on the selectivity estimation problem and the results show that the proposed method can accurately model the delicate relationship between accuracy and sample size and is portable to many distributions.

In summary, this paper makes following contributions.
(1)~Design a deep learning model for predictable error and response time for AQP in spatial data analysis.
(2)~Apply this model to the {\bf selectivity estimation} algorithm to solve two problems, sample size estimation and accuracy prediction.
(3)~Validate the model through experiments and publish the pre-trained model for wide use.

\section{Related Work}
\label{sec:related_works}

{\bf Approximate query processing:} AQP is a common method in many spatial data management systems. In AQP, the answer is estimated by executing the query on a small sample of the dataset, instead of scanning entire dataset. AQP is applied on several problems such as selectivity estimation, clustering, and spatial partitioning~\cite{siddique2019comparing}. For example, SpatialHadoop~\cite{eldawy2015spatial}, ScalaGiST~\cite{LCC+14}, Simba~\cite{XLY+16}, SATO~\cite{VAW14}  use a sample of the input dataset to compute the minimum bounding rectangles (MBRs) for their spatial partitioning operation. Sampling is also used to cluster very large datasets~\cite{bejarano2011sampling, yu2011sample}. Specially, sampling is the fundamental method for many selectivity estimation algorithms for spatial data~\cite{acharya1999selectivity}. The main idea of AQP is the trade-offs between query response time and accuracy as shown in Figure~\ref{fig:trade_off}. The common drawback of existing systems is the lack of a mechanism to choose a suitable sampling ratio to achieve a desired accuracy. For instance, SpatialHadoop just chooses a fixed $1\%$ sample of dataset to compute partition MBRs, which is not always the best choice. DeepSampling addresses this challenge by suggesting the minimum sampling ratio such that the desired accuracy could be achieved. For non-spatial data, BlinkDB~\cite{agarwal2013blinkdb} provides a bounded errors for standard relational queries. However, BlinkDB assumes the independence of data dimensions, which is not applicable for spatial data.

{\noindent \bf Deep learning and spatial data:} In recent years, the research community has witnessed the rapid growth of research projects in the intersection of big spatial data and machine learning~\cite{sabek2019machine}. One of the important research directions is scalable statistical inference systems for big spatial data analysis. For instance, TurboReg~\cite{sabek2018turboreg} is a scalable framework for building spatial logistic regression models. TurboReg is built on top of Markov Logic Network, which is able to predict the presence and absence of spatial phenomena in a geographical area with reasonable accuracy. DeepSPACE~\cite{vorona2019deepspace} is a deep learning-based approximate geospatial query processing engine. DeepSPACE utilize the learned data distribution to provide a quick response for spatial queries with reasonable accuracy. Both TurboReg and DeepSPACE hold the common drawback that they cannot guarantee a required precision of their answers. DeepSampling aims to overcome this issue by providing a prediction model such that the required precision is always met with a reasonable of sampling ratio budget.

\section{Selectivity Estimation with Predicted Error and Response Time}
\label{sec:method}

\begin{figure*}[t]
    \centering
    \subfigure[Existing AQP engine]{\includegraphics[scale=0.33]{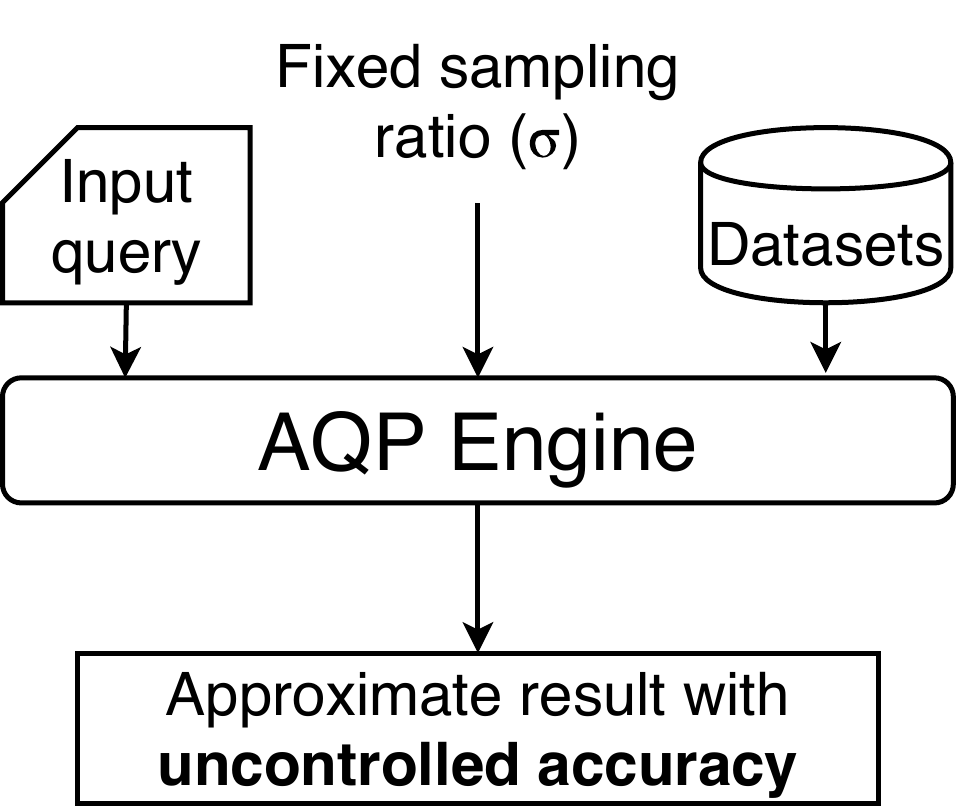}
    \label{fig:aqp:existing}}
    \hspace{12pt}
    \subfigure[Sample Ratio Estimation]{\includegraphics[scale=0.33]{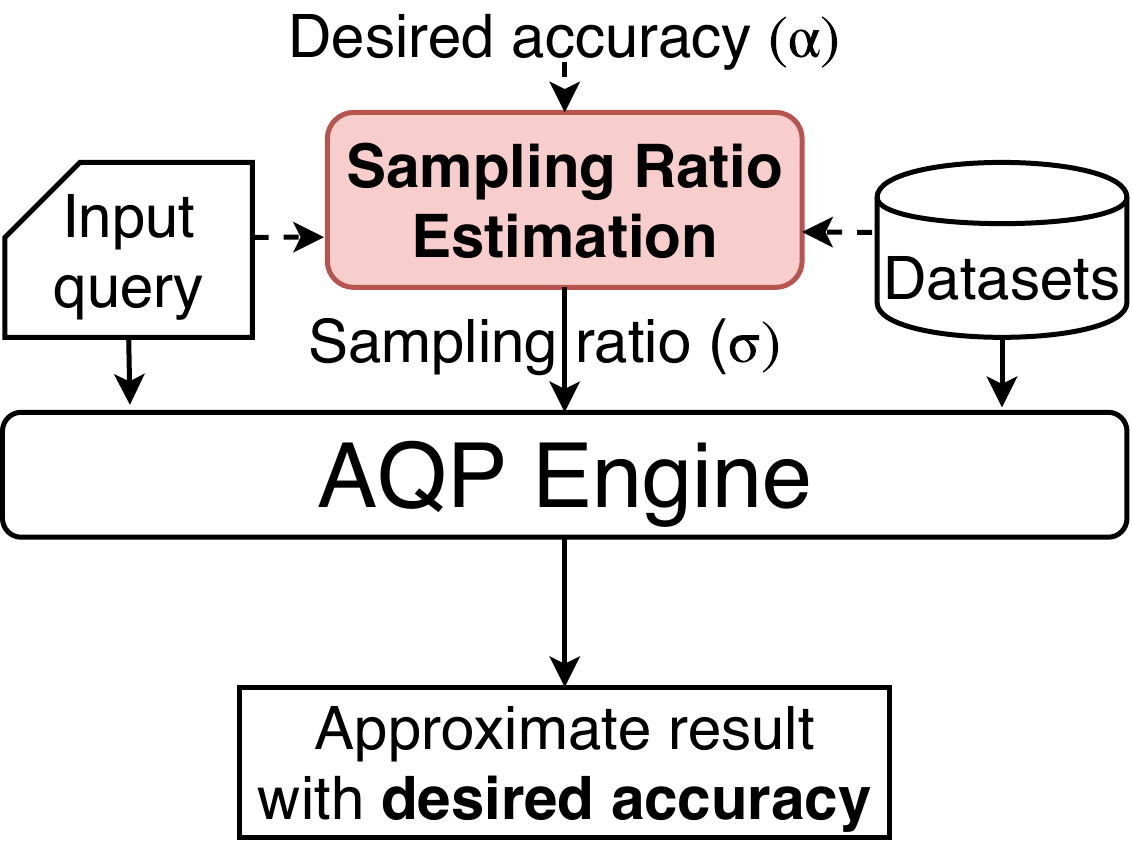}
    \label{fig:aqp:predict_sampling_ratio}}
    \hspace{12pt}
    \subfigure[Accuracy Prediction]{\includegraphics[scale=0.33]{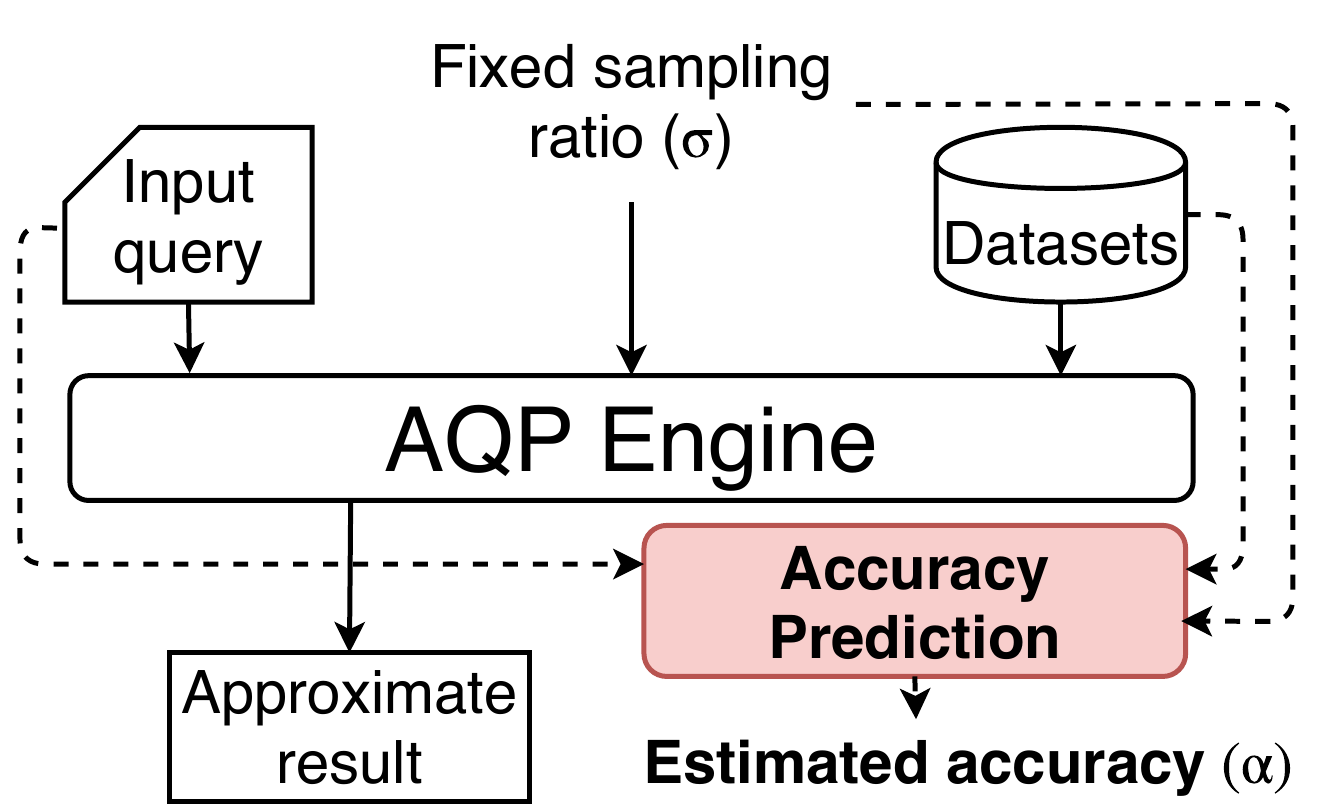}
    \label{fig:aqp:predict_accuracy}}
    \hspace{12pt}
    \caption{DeepSampling addresses critical problems on existing AQP systems}
    \label{fig:aqp}
\end{figure*}

\subsection{Problem definition}
\label{sec:method:problem_definition}

This paper focuses on the prediction model for the selectivity estimation problem but the proposed approach can be easily generalized to other problems such as K-means clustering or spatial partitioning. The goal is to find the relationship between accuracy and sample size and toward this goal we define two problems, {\em accuracy prediction} and {\em sample size estimation} which are both defined in this section. First, we will define the accuracy of an approximate answer in the selectivity estimation (SE) problem.

\begin{definition}[query accuracy]
In the SE problem, given an approximate answer $\pi$ and a ground truth $\Pi$ for query range $Q$, the accuracy of the approximate answer $\pi$ is
\begin{equation}
acc(\pi,\Pi)=max(0, 1 - |\Pi - \pi|/\Pi)
\end{equation}
\end{definition}

Based on this definition, we define the following two problems:

\textbf{Problem 1 (Sampling Ratio Estimation)}: \textit{Given a dataset $D$, a query range $Q$, and a desired accuracy $\alpha$, predict the minimum value of sampling ratio $\sigma$ such that $acc(\pi, \Pi) \ge \alpha$.}

\textbf{Problem 2 (Accuracy Prediction)}: \textit{Given a dataset $D$, a query range $Q$, and a sampling ratio $\sigma$, predict the accuracy $\alpha$ such that $|acc(\pi, \Pi) - \alpha|$ is minimized.}

Both problems are very important in approximate geospatial query processing. If we could address these problems, the existing spatial database systems could minimize the computation effort for sampling process while still achieving a desired accuracy for their answers. Figure~\ref{fig:aqp} shows how DeepSampling enhances performance of existing approximate query processing systems. Instead of fixing a sampling ratio as Figure~\ref{fig:aqp:existing}, an AQP engine can use Problem~1 to calculate a suggested minimal sampling size to achieve the used-desired accuracy as shown in Figure~\ref{fig:aqp:predict_sampling_ratio}. Conversely, if the system has a fixed sampling ratio, it can apply Problem~2 to estimate the result accuracy as shown in Figure~\ref{fig:aqp:predict_accuracy}.

\subsection{Prediction with mixed data sources}
\label{sec:method:prediction_mixed_data_sources}

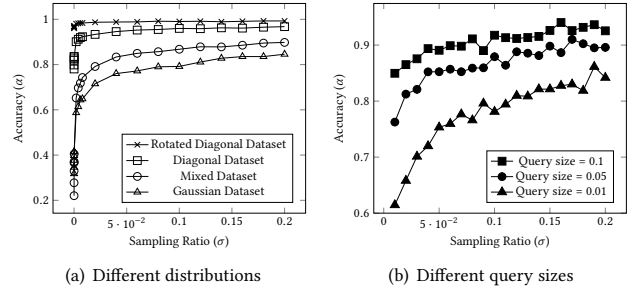
\begin{figure}[t]
    \centering
    \subfigure[Different distributions]{\pgfplotstableread{ 
Sampling_ratio	Combo_011.csv	Diagonal_004.csv	DiagonalRot_eny_005.csv	Gauss_006.csv
1.00E-05	0.2204725592	0.7805708184	0.9607314796	0.3177307608
2.00E-05	0.2775676215	0.7982824146	0.9623144388	0.3462828892
4.00E-05	0.3298375254	0.8189818789	0.9654365798	0.3767908447
6.00E-05	0.3658573501	0.8256309621	0.9680458247	0.3978194166
8.00E-05	0.3711388085	0.8340810421	0.9701971549	0.4141710582
0.0001	0.4048691802	0.8357873516	0.9707655221	0.4148718133
0.002	0.6523852706	0.9012584171	0.9805050743	0.5874976598
0.004	0.6968132661	0.9119325225	0.9822551776	0.6140761384
0.006	0.71800269	0.9194027365	0.9837937076	0.6437966079
0.008	0.7428587182	0.9231951383	0.9837916976	0.6488058653
0.02	0.791539106	0.9334306369	0.987069461	0.7152225651
0.04	0.8335530416	0.9458652703	0.9881439065	0.7602941984
0.06	0.8496616809	0.9525116884	0.98878604	0.7719267352
0.08	0.8570434041	0.95503867	0.9917390361	0.7900657984
0.1	0.8677607476	0.9598136594	0.9903646482	0.7918059298
0.12	0.8786080036	0.9586955159	0.9915445363	0.8111753773
0.14	0.8781721967	0.9624859525	0.9892804534	0.8280274537
0.16	0.8855366162	0.9611311389	0.991272486	0.835377541
0.18	0.8950178258	0.9652400228	0.9928872683	0.8354032168
0.2	0.8980590165	0.9676906016	0.9928417176	0.8457960848
}\testdata

\begin{tikzpicture}[scale=0.49]
\begin{axis}[
    legend style={at={(0.95,0.05)}, anchor=south east, font=\large},
    label style={font=\large},
    tick label style={font=\large},
    xlabel={Sampling Ratio ($\sigma$)},
    ylabel={Accuracy ($\alpha$)}],
    \addplot[mark=x, mark size=3]
        table[x index=0,y index=3]{\testdata};
    \addplot[mark=square, mark size=3]
        table[x index=0,y index=2]{\testdata};
    \addplot[mark=o, mark size=3]
        table[x index=0,y index=1]{\testdata};
    \addplot[mark=triangle, mark size=3]
        table[x index=0,y index=4]{\testdata};
    \legend{Rotated Diagonal Dataset, Diagonal Dataset, Mixed Dataset, Gaussian Dataset}
\end{axis}
\end{tikzpicture} \label{fig:relationship:datasets}}
    \subfigure[Different query sizes]{\pgfplotstableread{ 
Sampling_ratio	0.01	0.05	0.1
0.01	0.6145190659	0.7625529518	0.8496842703
0.02	0.6580425702	0.8124807919	0.8651213676
0.03	0.7011335864	0.821001513	0.8755588102
0.04	0.7198163275	0.8522804015	0.8937119501
0.05	0.753250135	0.852427442	0.8907679698
0.06	0.7596073455	0.8569808627	0.8989376543
0.07	0.77657932	0.8524341718	0.898042458
0.08	0.765964914	0.8589202475	0.9111172745
0.09	0.7959444772	0.8594624177	0.8901600421
0.1	0.7811567249	0.879251542	0.9176538863
0.11	0.7941734397	0.864056463	0.9132411195
0.12	0.8097676325	0.8880540173	0.9117847048
0.13	0.808656675	0.8855593124	0.9135055965
0.14	0.8213908672	0.8812756646	0.9155118333
0.15	0.8213480817	0.8981396324	0.9261403008
0.16	0.8275997809	0.886521505	0.9402807288
0.17	0.8300170838	0.9102232837	0.9258791956
0.18	0.8187293994	0.9024248702	0.9316960278
0.19	0.861390003	0.8950908517	0.936396206
0.2	0.8418880833	0.8959319257	0.9255009107
}\testdata

\pgfplotsset{ymin=0.6}
\begin{tikzpicture}[scale=0.49]
\begin{axis}[
    legend style={at={(0.95,0.05)}, anchor=south east, font=\large},
    label style={font=\large},
    tick label style={font=\large},
    xlabel={Sampling Ratio ($\sigma$)},
    ylabel={Accuracy ($\alpha$)}],
    \addplot[mark=square*, mark size=3]
        table[x index=0,y index=3]{\testdata};
    \addplot[mark=oplus*, mark size=3]
        table[x index=0,y index=2]{\testdata};
    \addplot[mark=triangle*, mark size=4]
        table[x index=0,y index=1]{\testdata};
    \legend{Query size = 0.1, Query size = 0.05, Query size = 0.01}
\end{axis}
\end{tikzpicture} \label{fig:relationship:selectivity}}
    \caption{How sampling ratio ($\sigma$) relates to accuracy ($\alpha$)}
    \label{fig:relationship}
\end{figure}

In general, we know that the accuracy ($\alpha$) of an approximate answer increases with the sampling ratio ($\sigma$). However, we show in this part that this relationship is more complex than that. Figure~\ref{fig:relationship} shows examples of how these two quantities are related to each other. First, Figure~\ref{fig:relationship:datasets} shows that this relationship highly depends on the dataset distribution. While for all distributions the sampling ratio and accuracy are highly correlated, the relationship is different for each dataset. For example, for the rotated diagonal dataset, the accuracy ranges from 96\% to 99\% for all sampling ratios while for the mixed distribution dataset, the accuracy ranges from 22\% to 90\%. Second, Figure~\ref{fig:relationship:selectivity} shows the relationship for different query sizes. This time, we see that the relationship highly depends on the query size as well.

These observations show how challenging the problem is. To build an accurate model, we need to take into account the input data distribution and the query size. For other problems, the query size could be replaced with other query parameters, e.g., the number of clusters for the K-means clustering problem, or the number of partitions for the spatial partitioning problem.

\subsection{DeepSampling architecture}
\label{sec:method:model_architecture}

\begin{figure}[t]
    \centering
    \includegraphics[width=0.5\linewidth]{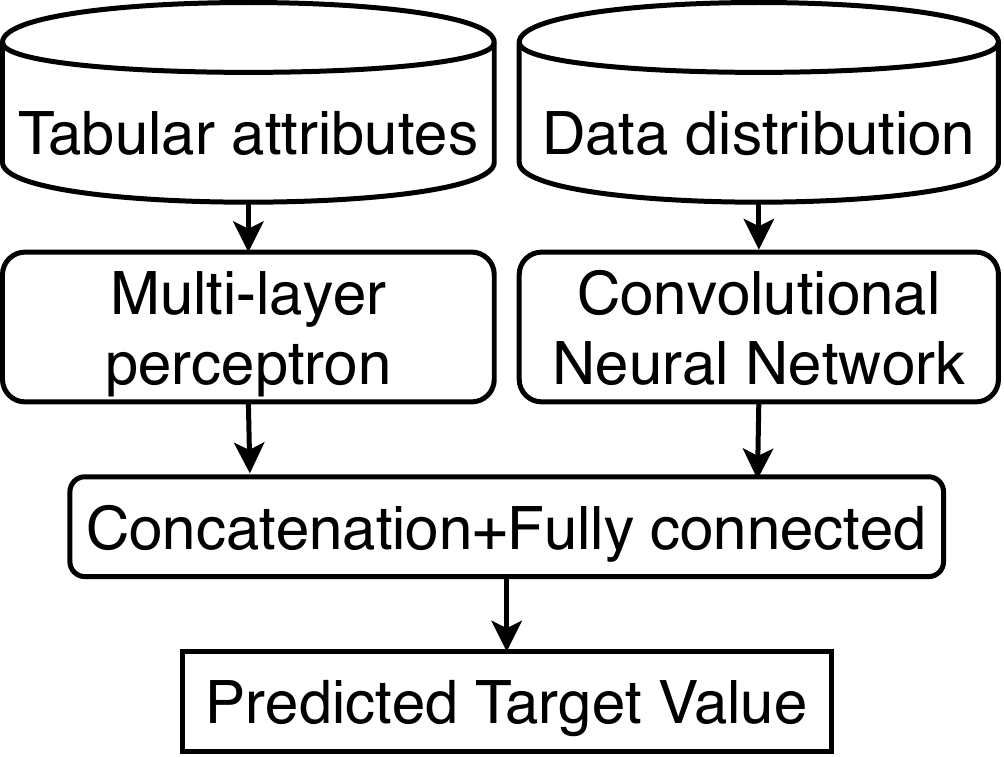}
    \caption{DeepSampling architecture}
    \label{fig:model_architecture}
\end{figure}

Figure~\ref{fig:model_architecture} shows an overview of the proposed architecture of the DeepSampling model. This architecture is used to solve both problems described earlier, {\em sampling ratio estimation} and {\em accuracy prediction}. To avoid repetition, we write between (parentheses) the changes that need to be made for the {\em accuracy prediction} problem.

To build an accurate and portable model that accounts for the query size and the data distribution, the proposed model takes two sets of inputs, {\em tabular data} and {\em data distribution}.

The \textbf{tabular input layer} consists of data taken from the processing logs which includes the query size ($q$), the sampling ratio ($\sigma$), and the resulting accuracy ($\alpha$). If we need to apply this architecture for other problems, then the query size will be replaced with other query parameters, e.g., number of clusters. Also, the accuracy will be calculated differently. This data is passed to a multi-layer perceptron (MLP) model. MLP is a feedforward neural network with at least three layers of nodes: an input layer, a hidden layer and an output layer. We chose MLP for tabular input since it can be used to learn complex mathematical models by regression analysis~\cite{cybenko1989approximation}.

The \textbf{data distribution input layer} catches the distribution of the input dataset. In this paper, we use a uniform histogram which is expected to accurately catch the dataset distribution if computed at a reasonable resolution. The histogram resolution is a system parameter that we study in the experiments section. Since this histogram is a $2$D matrix with spatial relationship between histogram bins, it is fed to a convolutional neural network (CNN) layer.

The \textbf{concatenation layer} combines the output of the MLP and CNN layers together and feed them to a fully connected (FC) layer. The final layer of FC is a single node with linear activation so that the model output is the predicted sampling ratio (or accuracy). The \textbf{loss function} of the final node provides a feedback on how accurate the predicted value is. Based on the problem definition in Section~\ref{sec:method:problem_definition}, we use mean absolute percentage error (MAPE) as the loss function which is the average absolute percentage error of actual value $A_t$ and forecast value $F_t$ for all training points $t\in[1,n]$ as shown in Equation~\ref{eqn:mean_absolute_percentage_error}.
MAPE is commonly used in regression models since it is very intuitive interpretation for relative errors.

\begin{equation}
    MAPE = \frac{1}{n}\sum_{t=1}^{n} \left|\frac{A_t - F_t}{A_t}\right|
    \label{eqn:mean_absolute_percentage_error}
\end{equation}

\section{Preliminary results}
\label{sec:experiments}

This section gives some preliminary results when applying the proposed approach to the selectivity estimation problem. In particular, we wanted to answer the following questions:
\begin{enumerate}
    \item How accurately does the model account for the data distribution and query size?
    \item Can the model solve both problems efficiently?
    \item Is the model portable enough so that we can test it on a new data distribution that was not in the training set?
\end{enumerate}

\subsection{Experimental setup}
\label{sec:experiments:setup}

\begin{table}[t]
\caption{Parameters for the selectivity estimation (SE) query}
\small
\begin{tabular}{|l|p{2.0in}|}
\hline
Parameter                 & Values  (Default)                                            \\ \hline
Dataset distribution      & Uniform, Gaussian, Diagonal, Sierpinski, Bit, Parcel, Mixed                   \\ \hline
Sampling ratio ($\sigma$) & 0.001, 0.0015,...,0.2                                        \\ \hline
Query size ($q$)          & 0.01,0.02,...,0.1.                                           \\ \hline
Histogram size ($h$)      & $1\times 1$ $\ldots$ ($16\times 16$) $\ldots$ $64\times 64$  \\ \hline
\end{tabular}
\label{tab:parameters}
\end{table}

We implement the proposed model in Figure~\ref{fig:model_architecture} using Keras~\cite{chollet2015keras}. The source code, training data and models are available at~\cite{gitdeepsampling}.

\textbf{Datasets:} We use both synthetic and real datasets in our experiments. We generated a total of $144$ synthetic datasets using the open-source spatial data generator~\cite{vu2019spatial}. The dataset distributions are listed in Table~\ref{tab:parameters} and the detailed distribution parameters are included in the source code~\cite{gitdeepsampling}. We also used two real datasets: {\tt OSM-Nodes}~\cite{UCRSTAR/OSM2015/all_nodes} and {\tt OSM-Lakes}~\cite{UCRSTAR/OSM2015/lakes}. The real datasets are only used for testing but never for training the model.

\textbf{Parameters:} In addition to the dataset distribution, we also vary the sampling ratio ($\sigma$), the query size ($q$), and the histogram size ($h$). The query size is the ratio between the area of the query rectangle and the area of the input minimum bounding rectangle (MBR). Our query workload consists of square queries centered at random locations in the input space. Table~\ref{tab:parameters} summarizes all the parameters that we vary in our experiments. In total, our generated dataset contains $54,720$ data points.

\textbf{Metrics:} We use mean absolute percentage error (MAPE) to evaluate the accuracy of a prediction model. The lower the value of MAPE, the better the model is.

\textbf{Baseline method:} We compare the proposed model to a linear regression(LR) model which takes the tabular input and predict a numeric output. The reason behind this choice is that we want to see how the dataset distribution input makes a difference to the baseline which only takes query attributes into account.

\subsection{Accuracy Prediction}
\label{sec:experiments:accuracy_prediction}

\begin{figure}[t]
    \centering
    \subfigure[Accuracy Prediction problem]{\pgfplotstableread{ 
distributions_size	ds_synthetic	ds_real	lr_synthetic	lr_real
1	0.3432856267	0.4158517576	1.631181488	2.083731096
2	0.2262081854	0.3131406279	1.391206891	1.846318966
3	0.2173757951	0.3036138576	1.282960123	1.710036939
4	0.08333487422	0.249509981	1.208179013	1.606974589
5	0.05779533028	0.1982525972	1.164174747	1.538930953
6	0.05714217179	0.1758762149	1.135640395	1.488314733
7	0.02864659893	0.1649758207	1.117862849	1.45341829
}\testdata

\pgfplotsset{ymin=0.0, ymax=2.2}
\begin{tikzpicture}[scale=0.48]
\begin{axis}[
    label style={font=\large},
    tick label style={font=\large},
    xtick=data,     
    xticklabels from table={\testdata}{distributions_size},
    legend style={at={(0.95,0.45)}, anchor=north east},
    xlabel={Number of training distributions},
    ylabel={Mean absolute percentage \textbf{error}}],
    \addplot[mark=triangle, mark size=4]
        table[x index=0,y index=1]{\testdata};
    \addplot[mark=triangle*, mark size=4]
        table[x index=0,y index=2]{\testdata};
    \addplot[mark=square, mark size=4]
        table[x index=0,y index=3]{\testdata};
    \addplot[mark=square*, mark size=4]
        table[x index=0,y index=4]{\testdata};
    \legend{DS-synthetic, DS-real, LR-synthetic, LR-real}
\end{axis}
\end{tikzpicture} \label{fig:experiment:accuracy_prediction}}
    \subfigure[Sampling Ratio Estimation problem]{\pgfplotstableread{ 
distributions_size	ds_synthetic	ds_real	lr_synthetic	lr_real
1	1.149445875	2.320015291	15.80074032	33.88739603
2	0.9797533823	1.397910977	7.683912878	14.94609209
3	0.8784878237	0.9301237684	7.491276552	13.11296829
4	0.8542443759	0.953151645	7.315879741	12.14997864
5	0.8024306192	0.9468109206	7.179288565	11.5445325
6	0.789596508	0.9324056359	7.156431331	11.09079064
7	0.7700513647	0.9416708613	7.060787319	10.74593437
}\testdata

\pgfplotsset{ymin=0.0, ymax=36.0}
\begin{tikzpicture}[scale=0.48]
\begin{axis}[
    xtick=data,     
    xticklabels from table={\testdata}{distributions_size},
    legend style={at={(0.95,0.95)}, anchor=north east, font=\large},
    label style={font=\large},
    tick label style={font=\large},
    xlabel={Number of training distributions},
    ylabel={Mean absolute percentage errors}],
    \addplot[mark=triangle, mark size=4]
        table[x index=0,y index=1]{\testdata};
    \addplot[mark=triangle*, mark size=4]
        table[x index=0,y index=2]{\testdata};
    \addplot[mark=square, mark size=4]
        table[x index=0,y index=3]{\testdata};
    \addplot[mark=square*, mark size=4]
        table[x index=0,y index=4]{\testdata};
    \legend{DS-synthetic, DS-real, LR-synthetic, LR-real}
\end{axis}
\end{tikzpicture} \label{fig:experiment:sampling_ratio_prediction}}
    \caption{Accuracy of DeepSampling and linear regression}
    \label{fig:performance_comparison}
\end{figure}
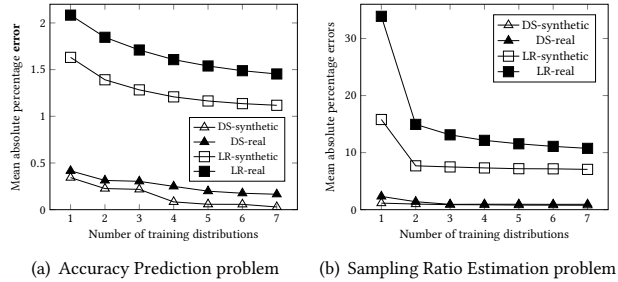

In the first experiment, we build a model to predict the average query accuracy, given the sampling ratio, query size and dataset histogram of size $16\times 16$. In particular, we use the synthetic datasets with $54,720$ data points described in Section~\ref{sec:experiments:setup} to train and test our proposed model. To observe how training data distribution affects the test accuracy, we organized the training data into different combinations of $1$ to $7$ distributions in Table~\ref{tab:parameters}. For each combination, we take a split of $75\%$ data points for training process. We test all the trained models with $25\%$ of the synthetic data points. We also test on $2800$ data points that we collected from SE queries on real {\tt OSM-Nodes} and {\tt OSM-Lakes} dataset. 

Figure~\ref{fig:experiment:accuracy_prediction} shows an interesting observation that the more data distributions we used for training process, the more accurate it is. This is expected since some simple distributions might not be able to capture important insights of test datasets. DeepSampling model is doing very well when we tested on both synthetic data and real data (MAPE is around $3\%$ and $16\%$). This shows the portability of the model. Even though the model was trained only on synthetic data, it still provided good results for the real dataset. In the future, we plan to add more synthetic data to make the model even more accurate with real data.
On the other hand, the linear regression baseline, due to its simplicity and the lack of data distribution, did not achieve a good accuracy. For the test on real dataset, its prediction is even more than $100\%$ beyond the actual mean accuracy value.

\subsection{Sampling Ratio Estimation}
\label{sec:experiments:sampling_ratio_prediction}

In this experiment, we build a model based on DeepSampling to predict sampling ratio, given a desired query accuracy, query size and dataset histogram of size $16\times 16$. We use the same set of training and testing split as mentioned in Section~\ref{sec:experiments:accuracy_prediction}. 

Table~\ref{fig:experiment:sampling_ratio_prediction} shows that DeepSampling is still doing better than the baseline when applied on both synthetic and real data. The errors are relatively higher than the accuracy prediction problem in Section~\ref{sec:experiments:accuracy_prediction}. The reason is that the range of the accuracy in the training set is narrow as compared to the range of sampling ratio. For example, in Figure~\ref{fig:relationship}, the accuracy in some cases stays above 95\% while the sampling ratio ranges from 0.1\% to 20\%. Nonetheless, the DeepSampling approach is consistently more accurate than the linear regression baseline. These results are consistent with existing work that found that the sampling ratio estimation problem is more difficult. For example, in BlinkDB~\cite{agarwal2013blinkdb} this problem is solved by simply choosing from a predefined set of points, sampling ratio and accuracy, and interpolating between them if needed.

\subsection{Effect of histogram resolution}
\label{sec:experiments:histogram_effect}

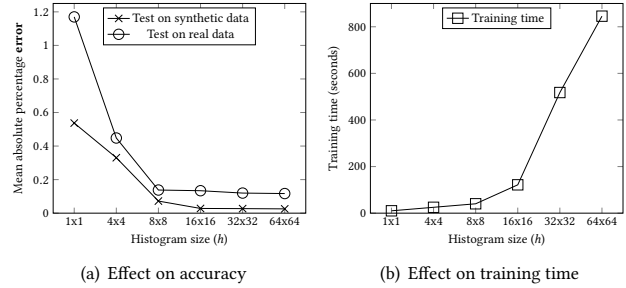
\begin{figure}[t]
    \centering
    \subfigure[Effect on accuracy]{
\pgfplotstableread{ 
size	test_on_synthetic	test_on_real	training_time
1x1	0.5370609293	1.169920564	10.1096189
4x4	0.3304102466	0.447953836	25.16631007
8x8	0.07206833862	0.1378744256	40.22784185
16x16	0.02811020943	0.1338336679	121.6015358
32x32	0.0270638747	0.1197349915	517.7928381
64x64	0.02559722496	0.1168719617	845.537791
}\testdata

\begin{tikzpicture}[scale=0.49]
\begin{axis}[
legend style={at={(0.5,0.98)}, anchor=north, font=\large},
ymin=0,         
ymax=1.25,
label style={font=\large},
tick label style={font=\large},
xtick=data,     
xticklabels from table={\testdata}{size} ,
xlabel={Histogram size ($h$)},
ylabel={Mean absolute percentage \textbf{error}}
]

\addplot[mark=x, mark size=4]
        table [x expr=\coordindex, y=test_on_synthetic, meta=size] {\testdata};\label{plot_test_on_synthetic}
\addplot[mark=o, mark size=4]
        table [x expr=\coordindex, y=test_on_real, meta=size] {\testdata};\label{plot_test_on_real}
        
\addlegendimage{/pgfplots/refstyle=plot_test_on_synthetic}
\addlegendimage{/pgfplots/refstyle=plot_test_on_real}

\legend{Test on synthetic data, Test on real data}

\end{axis}


   


\end{tikzpicture} \label{fig:histogram-resolution-accuracy}}
    \subfigure[Effect on training time]{
\pgfplotstableread{ 
size	test_on_synthetic	test_on_real	training_time
1x1	0.5370609293	1.169920564	10.1096189
4x4	0.3304102466	0.447953836	25.16631007
8x8	0.07206833862	0.1378744256	40.22784185
16x16	0.02811020943	0.1338336679	121.6015358
32x32	0.0270638747	0.1197349915	517.7928381
64x64	0.02559722496	0.1168719617	845.537791
}\testdata

\begin{tikzpicture}[scale=0.49]



\begin{axis}[
legend style={at={(0.5,0.98)}, anchor=north, font=\large},
ymin=0,         
ymax=900,
xtick=data,     
label style={font=\large},
tick label style={font=\large},
xticklabels from table={\testdata}{size} ,
xlabel={Histogram size ($h$)},
ylabel={Training time (seconds)}
]


\addplot[mark=square, mark size=4]
        table [x expr=\coordindex, y=training_time, meta=size] {\testdata}; 
   
\legend{Training time}

\end{axis}

\end{tikzpicture} \label{fig:histogram-resolution-training-time}}
    \caption{Effect of histogram resolution}
    \label{fig:histogram_effect}
\end{figure}

To choose a good histogram size, this experiment studies the trade-off between the model accuracy and training time as we vary the histogram size as depicted in Figure~\ref{fig:histogram_effect}. In this experiment, we vary the histogram resolution from $1\times 1$ (effectively no histogram) to $64\times 64$. Figure~\ref{fig:histogram-resolution-accuracy} shows the accuracy of the model when tested on both synthetic and real data as the histogram size increases. It is clear from this experiment that the histograms with higher resolutions carry more information that makes the model more accurate. However, the model stabilizes at $16\times 16$ where the histogram is accurate enough to catch the distributions in the training set.

Figure~\ref{fig:histogram-resolution-training-time} shows the total time of the training phase, i.e., the time until the model stabilizes. As expected, the model takes more time to train as the histogram resolution increases due to the large input that goes through the CNN model. From this experiment, we choose to set the histogram size to $16\times 16$ which gives a good accuracy in a reasonable time.

\section{Summary and future work}
\label{sec:conclusion}

In this paper, we introduced DeepSampling, a deep-learning-based system that provides predicted errors for approximate geospatial query processing. The proposed model combines the sampling ratio, the result accuracy, the query parameters, and the input data distribution. We carry some preliminary results when we apply DeepSampling to improve performance of selectivity estimation query. The results show that the proposed model can accurately compute the sampling ratio and accuracy for many synthetic and real distributions. In the future, we will apply the same model on other important approximate spatial problems such as K-means clustering and spatial partitioning. 

\bibliographystyle{ACM-Reference-Format}
  \bibliography{main}

\end{document}